 \documentclass[12pt,preprint]{aastex}

\newcommand{\sst}{{\it Spitzer Space Telescope}}

\slugcomment{To Appear in The Astrophysical Journal (Supplement Series),\\
\emph{Spitzer} Space Telescope special issue}

\shorttitle{Dust In Early-Type Galaxies}
\shortauthors{Pahre et al.}

\begin{document}

\title{Spatial Distribution of Warm Dust in Early-Type Galaxies.}

\author{Michael A. Pahre,
  M. L. N. Ashby,
  G. G. Fazio, and
  S. P. Willner}

\affil{Harvard-Smithsonian Center for Astrophysics, 60 Garden Street, Cambridge, MA  
  02138\email{mpahre, mashby, gfazio, swillner@cfa.harvard.edu} }

\begin{abstract}
Images taken with the IRAC instrument on the \sst\ show that the spatial
distribution of warm dust emission in lenticular galaxies is often 
organized into dynamically-stable structures strongly resembling spiral arms.
These galaxies have bulge-to-disk-ratios and colors for their stellar content
appropriate for their morphological classification.
Two of the three galaxies with warm dust detected at 8.0~\micron\ also show 
far-IR emission expected from that dust.
More importantly, the $[5.8]-[8.0]$ color of the dust emission matches the colors 
found for late-type, star-forming galaxies, as well as theoretical predictions
for PAH emission from dust grains.
The spatially resolved dust structures may be powerful indicators of the 
evolutionary history of the lenticular class of galaxies, either as a tracer
of on-going quiescent star formation or as a fossil record of a previous episode
of more active star formation.
\end{abstract}

\keywords{infrared: galaxies --- stars: formation --- dust, extinction --- ISM: lines and bands --- galaxies: fundamental parameters}

\section{Introduction}

While late-type spiral galaxies have long been known to contain an interstellar
medium in cold, warm, and hot phases, 
only during the last several decades was the same demonstrated for
at least some early-type galaxies.
Many early-type galaxies, for example, have significant quantities of cold gas in H~I
(Knapp et al. 1985; van Gorkom \& Schiminovich 1997) 
and hot gas seen at X-ray wavelengths (Kim et al. 1992).

The warm gas phase was detected in early-type galaxies
via its dust emission at far-IR
wavelengths with IRAS \citep{jura87,knapp89}
and at optical wavelengths from atomic gas emission \citep{caldwell84,phillips86}.
Later, ISO detected polycyclic aromatic hydrocarbon (PAH) lines in the mid-IR 
\citep{lu03,xilouris04}.
Of order half or more early-type galaxies exhibit either far-IR dust
or optical ionized gas emission, although the detection rate in the infrared could be lower
if the effects of an AGN are removed \citep{bregman98}.
While ionized gas regions are typically extended structures \citep{pogge93,goudfrooij94}, 
obscuration at optical wavelengths implies dust masses that are a small fraction of those 
implied by the infrared emission.
Ionized gas may be an incomplete tracer of the warm ISM because it 
will only appear in regions where there is sufficient
massive star formation to produce ionizing photons.
Current knowledge of the distribution of warm dust emission within early-type galaxies
indicates that it mostly follows the light distributions (e.g., Athey et al. 2002)
and hence is thought to arise from processes other than star formation such as AGN mass loss.
The exceptions to this statement are a few galaxies with somewhat extended
structures in $15$~\micron\ ISO images that have been interpreted as 
dust lanes \citep[e.g.,][]{xilouris04}.

This paper presents observations at $3 < \lambda < 10$~\micron\ taken with
the new \sst.
The mid-IR imaging capability of the Spitzer improves the spatial resolution
over previous missions by a factor of a few and the sensitivity by more than
an order of magnitude.
These new images show that several early-type galaxies have substantial structure in
the distribution of their warm dust as traced by PAH emission bands at $6.2$--$8.6$~\micron.





\section{Sample}

The six galaxies presented here are the first early-type galaxies observed from
a total sample of $\sim 100$ galaxies that will be observed with the \sst.
This 100 galaxy sample is, in turn, drawn from a ``parent''
sample of 
Ho, Filippenko, \& Sargent (1997),
which includes nearly 500 northern galaxies brighter than $B_T = 12.5$~mag.
Some details are given in a companion paper \cite{pahre04a}.
We consider the 100 galaxy sample to be ``representative'' of the magnitude-limited
parent sample, as the former was drawn to represent the full range of morphological types
and luminosities found in the latter.


\section{Observations}


The data were taken with the InfraRed Array Camera 
\citep[IRAC;][]{irac} on the {\it Spitzer Space Telescope} \cite{sstpaper}
during the first four IRAC campaigns of normal operations (2003 December -- 2004
March).  Six galaxies were imaged, each using five exposures of
either 12 or 30 seconds each on the source in all four detectors
(3.6, 4.5, 5.8, and 8.0~\micron).
The standard pipeline BCD (Basic Calibrated Data, versions 8.9 through 9.5)
products were used in the reductions.   
The first 5.8~\micron\ image taken for each galaxy
had an additional ``first-frame delta skydark'' subtracted from it, 
which was based on an independent dataset.
This additional dark image subtraction step corrects an instrumental artifact
that is not removed using the current pipeline processing library of sky dark images.
The bias level for the 5.8~\micron\ images was
significantly variable such that the median value of the background plus bias 
(which was iteratively determined via a sigma-clipping algorithm) had to be
subtracted from each individual frame before mosaicing.

A mosaic of each target in each bandpass was made using custom
software operating under the {\sc iraf} environment and using the
{\sc wcstools} package \cite{wcstools}.  Optical distortion was removed in the
process, and the images were subpixellated to $0.86$~arcsec, which is
a linear reduction by a factor of $\sqrt{2}$ in each dimension.  A
first iteration mosaic was constructed using a median combination of
the individual frames.  Each separate BCD frame was then compared
with the mosaic, and differences much larger than the amount expected from
the known noise were assumed to be cosmic rays or other artifacts.
Pixels affected were flagged as invalid, and the resulting
cleaned frames were then combined into the final mosaic. 
There are some residual instrumental artifacts remaining in the images, 
such as banding and muxbleed\footnote{See http://ssc.spitzer.caltech.edu/irac/dh/ .}, 
but these are masked out for the surface photometry described below.
The resulting 3.6~\micron\ images are shown in Figures~\ref{fig1} and \ref{fig2}.

\placefigure{\fig1}
\placefigure{\fig2}

Quantitative data in this paper are given in instrumental magnitudes
relative to Vega.  The existing calibration of IRAC data is based on
point source observations \citep{irac}.
Extended sources will have
aperture corrections different from those for point sources, and the
correct values are still unknown.  
For this work, we base aperture corrections on 
the elliptical galaxy \dataset[ads/sa.spitzer#0004431104]{NGC~777}, 
which shows the least evidence for any dust emission.\footnote{We 
assume a pure stellar spectrum matching that of an M0~III star.
Similar results would be obtained if the assumption were made for
galaxies \dataset[ads/sa.spitzer#0004478976]{NGC~5077} or 
\dataset[ads/sa.spitzer#0004490496]{NGC~5813}.
Since the near-infrared colors of old stellar populations match
M~giants \cite{frogel78}, we have chosen to assume an M~giant 
SED for the purpose of calibrating the mid-IR aperture corrections.}
The resulting corrections, representing the ratio of point source
flux in a 12.2~arcsec aperture to that in an infinite aperture, are
$(0.09,0.05,0.17,0.31)$~mag for wavelengths $(3.6,4.5,5.8,8.0)$~\micron.\footnote{An 
independent estimate of the extended source aperture correction was
derived from the wings of the PSF from an observation of the bright star Fomalhaut.
The results from NGC~777 derived here agree within
$0.03$~mag with preliminary estimates based on the flux at $r >
12.2$~arcsec from the Fomalhaut observations---except for
5.8~\micron, where the galaxy magnitudes implied by the Fomalhaut
data are $\sim 0.15$~mag fainter.  
None of our main conclusions depends on the uncertain 5.8~\micron\ aperture correction.}
If NGC~777 does have dust emission after all, then the other galaxies will have
correspondingly larger amounts of dust emission than derived here.
The absolute calibration is based on
Vega flux densities of
$(277.5,179.5,116.6,63.1)$~Jy at effective wavelengths of (3.55,
4.49, 5.66, and 7.84~\micron\ in the four IRAC bands.

\section{Analysis}

\begin{figure}
\epsscale{0.80}
\plotone{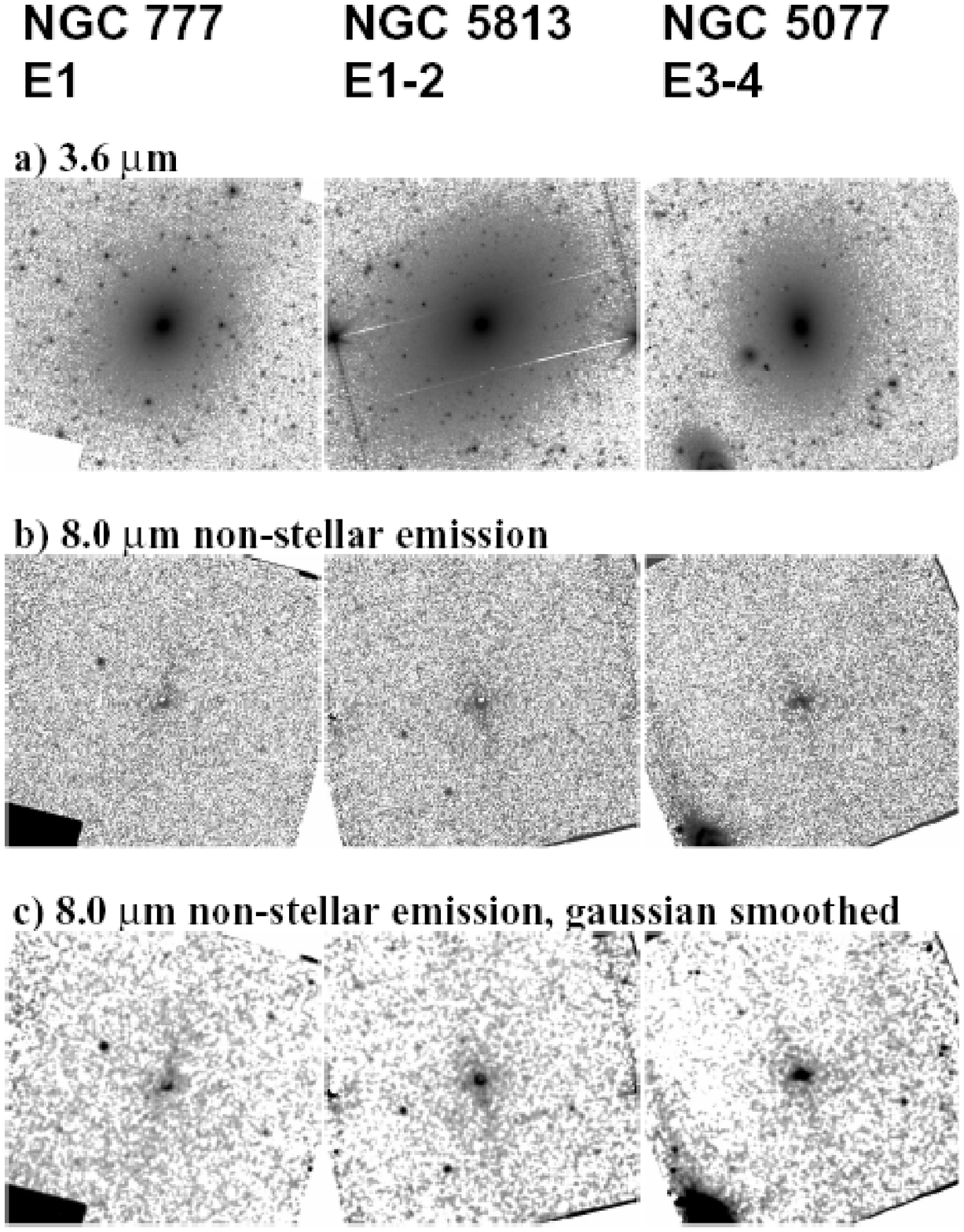}
\caption{\footnotesize
PLATE 1.
Greyscale images of the three elliptical galaxies taken with the IRAC instrument on \sst.
All images are $\sim 5$~arcmin on a side with North up and East to the left.
Images of NGC~777 are shown on the left-hand side column, NGC~5813 in the middle column, 
and NGC~5077 in the right-hand column.
(a) Images at $\lambda = 3.6$~\micron, which represents the stellar emission from each galaxy.
(b) Non-stellar emission images at $\lambda = 8.0$~\micron, which was created by subtracting
a scaled-down version of the 3.6~\micron\ image from the 8.0~\micron\ image (see text in 
Section \S~3).  
(c) Smoothed versions of (b), using a gaussian kernel of $\sigma = 1.5$~pixel ($1.3$~arcsec),
in order to bring out any faint features.
All images are shown with a ``square-root'' function for the stretch; the minimum (maximum)
displayed values are 0~(5)~MJy/sr for (a) and (b), and 0~(0.5)~MJy/sr for (c).
These three galaxies show little or no evidence for emission from warm dust at 8.0~\micron, 
although NGC~5077 does seem to show evidence for a weak AGN.
%
\label{fig1} }
\end{figure}

For each galaxy, the $[3.6]-[4.5]$ color is nearly constant with
radius and is consistent with stellar photosphere emission. Dust
emission is likely to be important at longer wavelengths.
The stellar emission was subtracted from the $5.8$ and $8.0$~\micron\
images using the average of the $3.6$ and $4.5$~\micron\ images, each 
suitably scaled to match the theoretical colors of M0~III stars: 
$[3.6]-[4.5]=-0.15, [4.5]-[5.8] = +0.11, [5.8]-[8.0] = +0.04$~mag 
(M. Cohen and T. Megeath, private comm.).\footnote{The blue color
$[3.6]-[4.5]=-0.15$ is most likely due to CO absorption in the 4.5~\micron\ bandpass
for the cool giant stars that dominate the light at these wavelengths (M. Cohen, private comm.).}
For example, the 3.6~\micron\ image (in units of MJy/sr) would be multiplied
by a factor of $10^{-0.04/2.5} \times 10^{(0.09-0.17)/2.5} \times {116.6 \over 277.5}$
in order to model the 5.8~\micron\ image flux, where the first term accounts for the
assumed stellar color (Vega-relative), the second term has the aperture corrections for the two
wavelengths, and the third term is the ratio of the flux of Vega in the two bandpasses.
The residual 8.0~\micron\ images are referred to as
``non-stellar'' emission\footnote{Some
authors in the literature refer to images such as these as ``dust images,''
but we choose to use the term ``non-stellar images'' to reinforce the
point that AGN emission could be present.} and are shown in Figures~\ref{fig1} and \ref{fig2}.
These non-stellar emission images were also smoothed using a gaussian kernal
of $\sigma = 1.5$~pixel ($1.3$~arcsec) in order to highlight any faint, extended emission
features.
The smoothed images are also shown in Figures~\ref{fig1} and \ref{fig2}.


Surface photometry was measured on the galaxies using the IRAF task
{\sc ELLIPSE}, following the methodology of Pahre
(1999).  Point sources were identified and masked using
the 2MASS point source catalog (PSC) while excluding the source at the galaxy's location;
additional, fainter point sources were identified and masked by eye
in the $3.6$~\micron\ image.  The surface photometry was first measured on the $3.6$~\micron\ 
image, then the flux at each isophote was measured for the
other wavelengths using the isophotal shapes defined from the $3.6$~\micron\ 
image.  This ensures that the colors are measured for identical
regions of the galaxy in all images.  
Two surface brightness profiles were calculated for each galaxy:
one allowing ellipticity to vary with radius, and another using circular apertures.
Corrections for flux lost due to
the PSF were applied to the surface brightness and aperture magnitude profiles
in the manner described by Pahre (1999); while this approach is only 
strictly accurate for bulge-like light distributions, the profiles are
sampled and fitted at such large radii relative to the PSF that the
corrections themselves are very small.

One-dimensional models of $r^{1/4}$ bulge plus exponential disk
were fit to the isophotal surface brightness profiles
and the circular aperture magnitude curves of growth.
The results of the two methods were averaged, and half of the difference
taken as the uncertainty.
Results are tabulated in Pahre et al. (2004).
The images and profiles were inspected to determine robust outer fitting radii;
the value of $5/3$ times the 2MASS isophotal radius $\mu_{K_{20}}$  was found to
offer a good compromise of sampling most of the galaxy light and providing for
robust results independent of outer fitting radius while at the same time
minimizing the effects of sky subtraction.






\section{Spatially Resolved Dust Emission in Early-Type Galaxies}

\begin{figure}
\epsscale{0.80}
\plotone{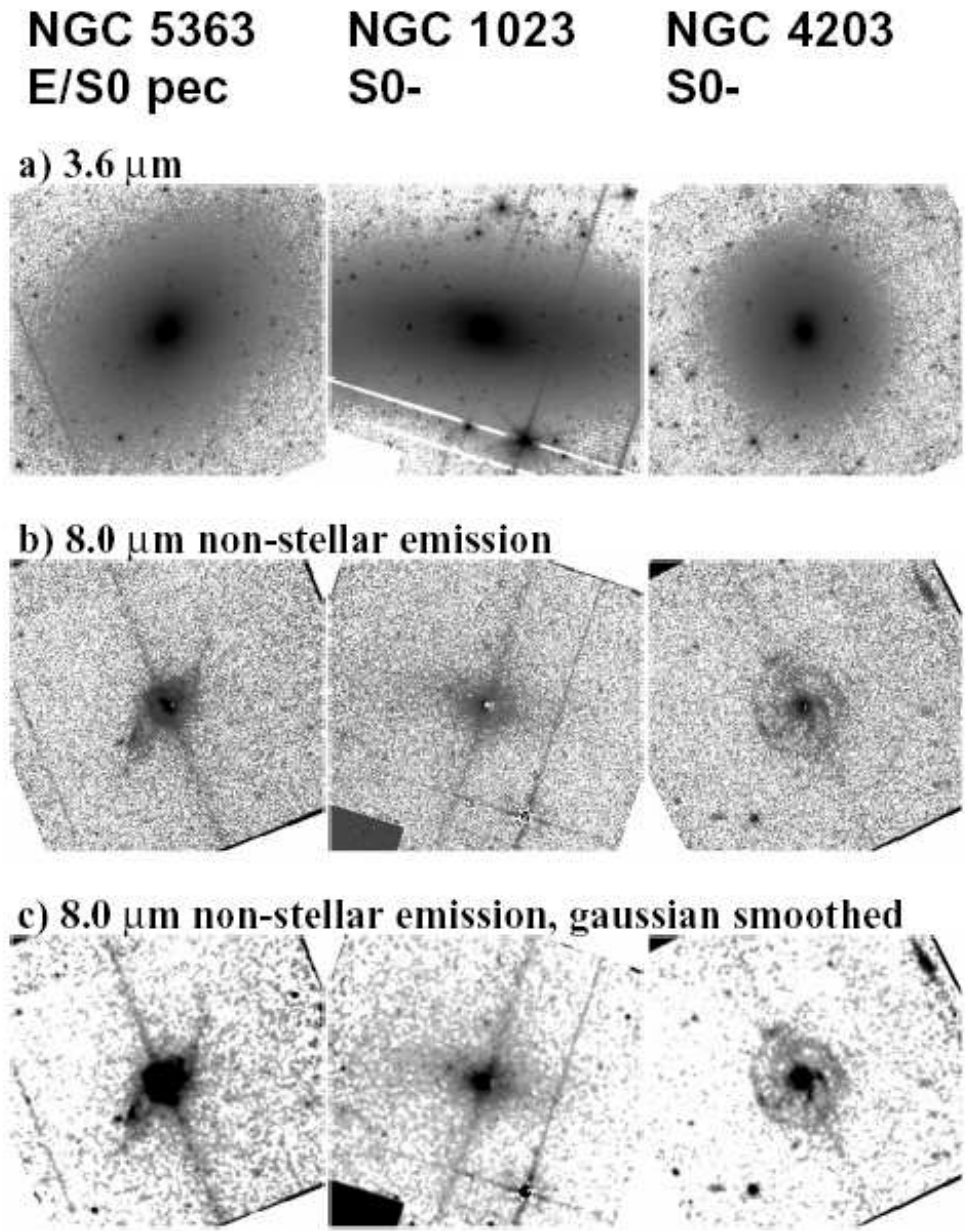}
\caption{
PLATE 2.
Greyscale images of the three lenticular galaxies taken with the IRAC instrument on \sst.
The panels are otherwise the same as in Figure~1.
Images of NGC~5363 are shown on the left-hand side column, NGC~1023 in the middle column, 
and NGC~4203 in the right-hand column.
These three lenticular galaxies show evidence for substantial non-stellar emission
at 8.0~\micron, which is presumed to originate from the PAH lines at 7.7 and 8.2~\micron.
This dust emission appears as spiral arm-like structures for
NGC~5363 (apparently inclined) and NGC~4203 (roughly face-on).
\label{fig2} }
\end{figure}

The three galaxies classified as E/S0 or S0 (NGC~1023, 4203, and 5363) show substantial
residual, non-stellar emission in the $8.0$~\micron\ band (Fig.~\ref{fig2}c),
while the three classified as E do not (Fig.~\ref{fig1}c).
The dust exhibits different morphologies in each galaxy.

In the case of S0- galaxy \dataset[ads/sa.spitzer#0004457216]{NGC~4203}, the warm dust emission appears 
to be organized into several spiral arms emanating out from the 
nuclear region.
Some evidence of nuclear dust obscuration has been found for NGC~4203 using optical colors 
on HST \cite{erwin03}, but the optical images do not show the wealth
of spiral arm structures at larger radii in the $8.0$~\micron\ image.

In the E/S0 galaxy \dataset[ads/sa.spitzer#0004484352]{NGC~5363}, the emission appears as a compact disk
with arms and a possible bar-like structure projected against the disk.
The galaxy shows additional, fainter, arm-like emission that extends farther 
out roughly along the major axis of galaxy.
This galaxy has been known for some time to exhibit dust obscuration \cite{gallagher86}.
Mid-IR images taken with ISO showed 
dust emission, slightly elongated
along the galaxy's minor axis (Xilouris et al. 2004), but their resolution and sensitivity
were insufficient to see the detailed distribution of the dust shown in Figure~\ref{fig2}c.

For S0- galaxy \dataset[ads/sa.spitzer#0004432640]{NGC~1023}, the dust emission is faint and 
smoothly distributed in a direction that roughly corresponds to
the major axis for the inner part of the galaxy.

Two of the three lenticular galaxies (NGC~4203 and NGC~5363), as well as
E3-4 galaxy \dataset[ads/sa.spitzer#0004478976]{NGC~5077}, show evidence 
for redder nuclear colors than their inner bulges, which is the mid-IR 
signature of the presence of an AGN.
All three have been argued to have LINER nuclei, while
NGC~5077 may instead harbor a Sy nucleus (Ho et al. 1997). 


While dust emission in early-type galaxies has been
known from previous IR space missions, Figure~\ref{fig2} provides the first
evidence that its spatial distribution often resembles
spiral arms---a hallmark of star formation in a quiescent disk.
Some dwarf elliptical galaxies have recently been shown also to
exhibit structures like these at optical wavelengths
(Barazza et al. 2002; Graham et al. 2003).

\begin{figure}
\epsscale{0.80}
\plotone{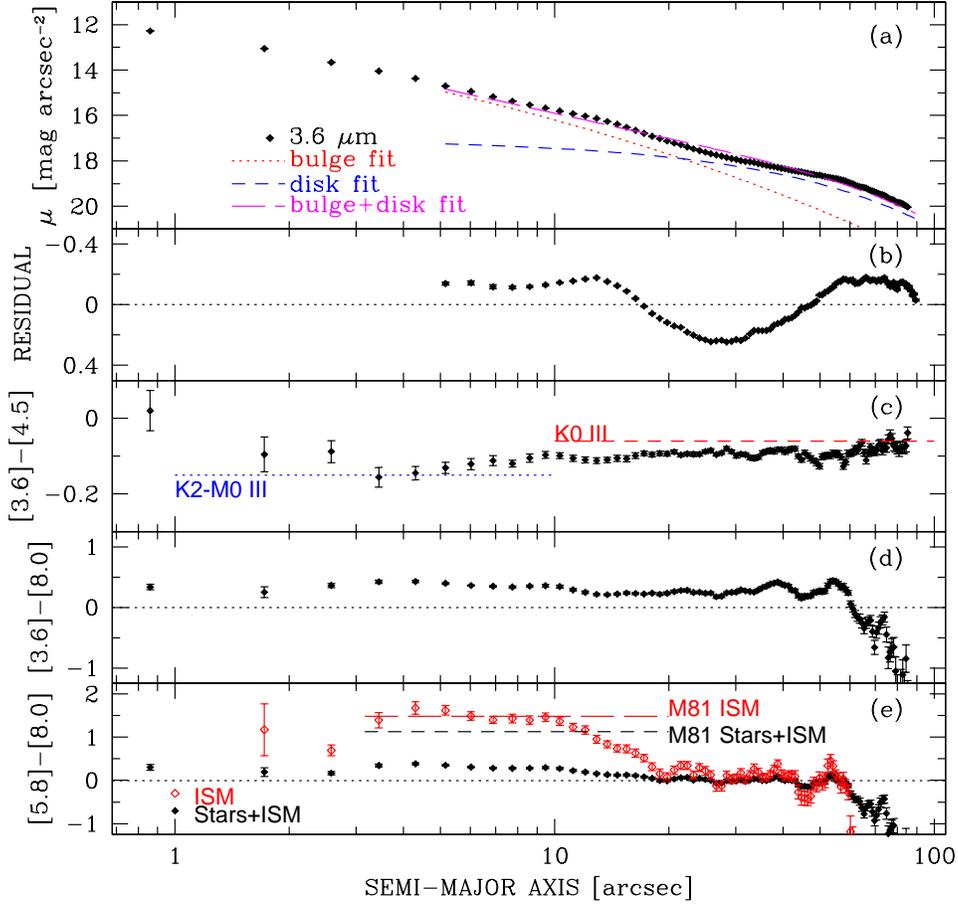}
\caption{ 
Surface brightness and color profiles for NGC~4203.
(a) The SB profile at $3.6$~\micron\ [black diamonds] is well-fit by a combined [magenta long dashed line]
bulge [red dotted line] and disk [blue short dash line] model of its light distribution with $B/D = 1.4$ (average
for the two methods; see text), typical
for S0 galaxies (Simien \& de Vaucouleurs 1986).
(b) Residuals of that model fit; this galaxy has the largest residuals in the sample.
The residuals at $r < 20$~arcsec could be indicative of faint underlying spiral structure
in the stellar light distribution.
(c) The $[3.6]-[4.5]$ color, which is a direct indicator of the mean spectral type
of the composite stellar population, shows a small reddening trend from the inner bulge
to the outer disk region.  This color change is comparable to small changes in stellar
spectral type among K giant stars from $-0.15$~mag (K2--M0 III) to $-0.06$~mag (K0 III).
(d) The $[3.6]-[8.0]$ color is modestly red, indicating the presence of warm dust emission
in the $8.0$~\micron\ band.
(e) The $[5.8]-[8.0]$ color for the total light ``Stars+ISM'' (black filled symbols) 
shows only a modest amount of emission (non-zero positive value),
because stellar light dominates over the dust in these bands.
But the $[5.8]-[8.0]$ color of the dust (red open symbols, ``ISM''), which has had the
stellar light subtracted from both bandpasses, is quite red,
matching the value found for the ISM for star-forming galaxy M~81 \cite{willner04}.
Similar results are also found for the photometric properties of the dust in NGC~1023 and
NGC~5363.
\label{fig3} }
\end{figure}

Mis-classification is not likely to explain the three S0 galaxies containing
warm dust.
Galaxy bulge-to-disk ratio correlates well with morphological type \cite{vdb76}.
The three galaxies NGC~1023, 4203, and 5363 have 
$B/D = 2.7 \pm 1.8, 1.4 \pm 0.5,$ and $4.7 \pm 1.9$, respectively,
measured at $3.6$~\micron\ \cite{pahre04a}.
Not only are these all reasonable values of $B/D$ for the lenticular class
\cite{simien86}, they are actually somewhat higher than the value of $1.35$
found for optical wavelengths, since the bulge light is generally redder than the disk.
The $3.6$~\micron\ surface brightness profile and its bulge plus disk fit are plotted
in Figure~\ref{fig3} for the worst fit in the sample (NGC~4203).
By comparison, the $B/D$ ratios for the elliptical galaxies NGC~777, 5077, and 5813 are
$14 \pm 3, 14 \pm 10,$ and $0.9 \pm 0.1$, respectively.  This demonstrates that the first two are
confidently fit by de~Vaucouleurs-type light distributions, while the latter might be
better classified as a lenticular.

\placefigure{\fig3}

Variations in stellar content cannot explain the observed $8.0$~\micron\ emission.
The color profile in Figure~\ref{fig3} shows that $[3.6]-[8.0]$ color is between $+0.2$ and $+0.5$~mag
throughout the extent of the disk.
This color is too red to be consistent with any stellar type or population,
and there is no reason to believe that there are many magnitudes
of visual extinction internal to any of these three galaxies.
In fact, the color profile in $[3.6]-[4.5]$ matches the colors of late K or M giant 
throughout all the galaxies, as is shown in Figure~\ref{fig3} for NGC~4203, 
except in regions close to the active nuclei.

The inferred colors of the non-stellar emission match what is found in galaxies
that show star formation.
Figure~\ref{fig3} shows that the inner region of NGC~4203
at $3 < a < 11$~arcsec has a non-stellar emission color of 
$[5.8]-[8.0] \sim 1.5 \pm 0.1$~mag, in good agreement with
the value of $1.48$~mag found for M~81 \cite{willner04}.
A similar value of the non-stellar color $[5.8]-[8.0]$ is found for the
dust emission region of NGC~1023, while NGC~5363 shows a 
larger color $\sim 2.0$~mag.
There is approximately $0.1$ to $0.2$~mag systematic uncertainty in the 
$5.8$~\micron\ aperture correction from $r=10$~pixel to $\infty$, which would act in the
direction to make $[5.8]-[8.0]$ more positive.
The PAH emission models of Li \& Draine (2001) predict that the IRAC color
should be $[5.8]-[8.0] \sim 2.06$~mag for a large range of five orders of
magnitude in incident UV radiation.
Thus, the color of the dust emission found in these early-type galaxies,
star-forming M~81, and a theoretical model of aromatic features from dust
grains, all have broadly consistent mid-IR SEDs.
We caution against interpreting the dust colors in themselves as strong evidence for star formation
in these early-type galaxies, however, because the model of Li \& Draine 
shows that such flux ratios occur independent of the properties of the
UV radiation field.
On the other hand, the emission is inconsistent with the $\lambda > 9$~\micron\ emission from AGB
mass loss seen in some nearby elliptical galaxies (Athey et al. 2002)---which would enter at the
long wavelength end of the 8.0~\micron\ filter bandpass---because the same warm dust emission
features are seen at 5.8~\micron, albeit at lower S/N.

\begin{figure}
\plotone{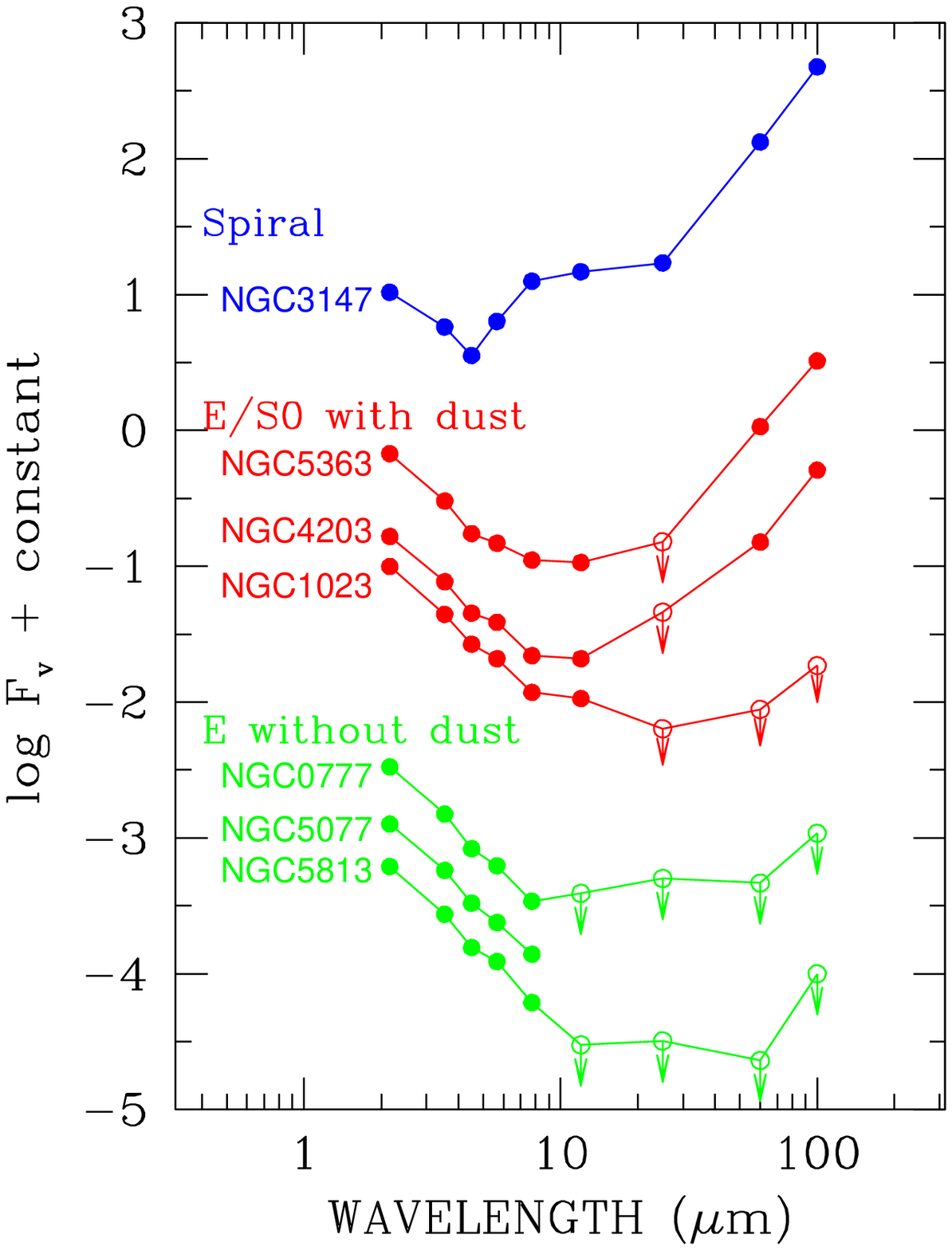}
\caption{ 
Infrared SEDs of the early-type galaxies in this paper.
The $2.2$~\micron\ data are taken from 2MASS, the $3 < \lambda < 10$~\micron\ data are
from IRAC (this paper), and the $12 < \lambda < 100$~\micron\ data are from IRAS
\cite{soifer89,irasfsc,knapp94}.
Upper limits in the IRAS catalog are plotted as open symbols with arrows.
The two galaxies with dust structures at $8.0$~\micron\ show clear evidence of
warm dust in the far-IR, while the third galaxy (NGC~1023), which appears to have
a smooth distribution of dust at $8.0$~\micron, does not.
The three galaxies without signs of dust emission at $8.0$~\micron\ all show no
evidence for warm dust in the far-IR.
A star-forming, spiral galaxy (NGC~3147) is plotted for comparison.
\label{fig4} }
\end{figure}

Galaxies with optical brightness similar to these have been studied with
IRAS in order to detect (or place strong limits on) the presence of warm dust.
A reasonable prediction would be that the three early-type galaxies 
with dust emission at $8.0$~\micron\ should also exhibit far-IR emission from that dust,
while those without $8.0$~\micron\ dust emission would show only upper limits in the far-IR.
The spectral energy distributions (SEDs) for the galaxies for
$2 < \lambda < 100$~\micron\ are plotted in Figure~\ref{fig4}.
All but one of the six early-type galaxies follow the expectations.
The exception is NGC~1023, whose stringent limits in the far-IR seem 
to contradict the presence of extended $8.0$~\micron\ warm dust emission.
This result for NGC~1023 cannot be explained by different detection limits (in Jy);
in fact, a far-IR SED shape similar to NGC~4203 or 5363 can be ruled out for NGC~1023,
as is apparent in Figure~\ref{fig4}.

\placefigure{\fig4}

The nature of the infrared dust emission from early-type galaxies remains
controversial.
On the one hand, Devereux \& Young (1990) and Devereux \& Hameed (1997) argue that
young stars are producing virtually all of the far-IR emission, and hence the far-IR
emission is a direct tracer of the star formation rate.
On the other hand, Kennicutt (1998) points out that many early-type galaxies show
no independent evidence for the presence of star formation other than the far-IR
emission, thereby suggesting that it may be an emission more similar to cool Galactic
cirrus clouds.
While the present data do not discriminate between these two origins for the infrared
emission, a multi-wavelength approach combining observations of ionized gas, warm dust,
and UV continuum emission could address it.
We defer this to a future contribution.

\section{Summary}

Images at $3 < \lambda < 10$~\micron\ taken with the IRAC instrument on \sst\ demonstrate:
\begin{enumerate}
\item Three of six early-type galaxies observed exhibit dust emission that is organized into
  spiral arm or inner disk-like structures,
\item The bulge-to-disk ratios of the galaxies support their classification as
  early-types,
\item The $[5.8]-[8.0]$ color of the dust emission matches that for dust in actively 
  star-forming galaxies and theoretical models of PAH emission, and
\item Two out of three galaxies that show $8$~\micron\ dust emission also show far-IR
  emission.
\end{enumerate}

\acknowledgments

M.A.P. acknowledges NASA/LTSA grant \# NAG5-10777.
The IRAC GTO program is supported by  JPL Contract \# 1256790.
The IRAC instrument was developed under JPL Contract \# 960541.
This work is based on observations made with the \sst, which is
operated by the Jet Propulsion Laboratory, California Institute of
Technology under NASA contract 1407. 

Facilities:  \facility{Spitzer(IRAC)}, \facility{2MASS}, \facility{IRAS}.

\clearpage

\clearpage
%

\end{document}